# Mathématiques et Mécanique à Strasbourg entre 1871 et 1939

## D. HUILIER[a]

a. Université de Strasbourg, CNRS, ICube, Strasbourg, France huilier@unistra.fr

## Résumé

*L'histoire de l'Université de Strasbourg, humaniste dès le début, est complexe, ce d'autant plus qu'elle a été ballotée entre possession prussienne, germanique et française. Si l'on étudie en particulier le domaine des mathématiques à partir de 1871, domaine qui verra naître progressivement une science dérivée des mathématiques appliquées, à savoir la « mécanique », force est de constater que les dirigeants politiques à travers cette trajectoire chaotique ont toujours nommés des scientifiques de valeurs. En mathématiques, pendant la domination prussienne (1871-1918), il s'agissait surtout d'en faire un deuxième pôle digne de Göttingen, puis pendant les débuts de la repossession française, une sorte d'antichambre pour les nominations de prestige à la Faculté des Sciences de Paris, la Sorbonne, l'Académie des Sciences ou encore l'ENS. La présente communication se propose de jeter un œil, par le biais de l'histoire, sur le développement des, et l'importance que prendront, les mathématiques à Strasbourg, en lien avec ce qui deviendront le futur Institut de Mécanique des Fluides et des Solides en 2000 et l'Institut de Recherche en Mathématique Avancée (IRMA).*

## Abstract

*The University of Strasbourg, fundamentally humanistic since its creation, has a complicated history, being sometimes German, sometimes French through the ages. If one focus, from 1871, on the area of mathematics, we can identify two periods. Political goals led to send high-level scientists, sometimes rather young, to develop theoretical, later applied mathematics, leading to a novel area, mechanics. The first period, under Prussian influence until 1918, will see the emergence of an Institute of mathematics such it was existing in Göttingen; the second, dominated by the French influence, will make Strasbourg an anteroom to positions at the prestigious Faculty of Sciences in Paris, the Sorbonne, the Academy of Sciences or the ENS. Present communication will give an inside, through its history, of the development of Mathematics in Strasbourg, prelude to emerging Mechanics and the creation of what will become the Institut de Mécanique des Fluides et des Solides in 2000 and IRMA, the Research Institute of Advanced Mathematics..*

**Mots clefs :** Histoire des mathématiques, période 1919-1939, Strasbourg, Mécanique théorique

## 1    Introduction

Si l'Université de Strasbourg peut être fière d'être un pôle attractif en Mathématiques pures depuis toujours, elle a aussi le privilège d'être le berceau d'une Mécanique naissante de par la nomination à la Faculté/Institut de Mathématiques de personnalités qui marqueront notamment le développement de la Mécanique Théorique et Appliquée, puis de la Mécanique des Fluides en France. L'Université de Strasbourg a un passé mouvementé, souvent sous influence germanique [1,2]. Si l'on se restreint aux mathématiques pures, physique mathématique et mathématiques appliquées, ces deux derniers termes convenant plus aux futurs mécaniciens, Strasbourg, sous domination allemande, et son Institut de Mathématiques verront passer des chercheurs éminents en mathématiques, Elwin Bruno Christoffel (1872-1894)*, Heinrich Weber (1895-1919) et son élève Joseph Wellstein (1904-1919), Ludwig Maurer (1886), Adolf Krazer (1872-1902), élève de Friedrich Emil Prym, lequel refusera un poste à Strasbourg, Maximilien Simon (1871-1912), élève de Karl Weierstrass et Ernst Kummer, Paul Epstein (1895-1918). Mais également des mathématiciens attirés par les applications à la physique et la mécanique tels que Theodor Reye (1872-1909), co-fondateur de l'Institut de Mathématiques, en fait du « Mathematisches Seminar » avec Christoffel en 1872, ou encore Richard von Mises (1909-1919), l'un des fondateurs d'une mécanique théorique « moderne » pluridisciplinaire. Redevenue française en 1918, l'Université de Strasbourg devra remercier les brillants chercheurs allemands encore actifs. Fortement soutenue par Raymond Poincaré, elle va retrouver rapidement son prestige d'Université française par la nomination, entre autres, de brillants mathématiciens tels que Paul Flamant, Maurice Fréchet, Georges Valiron, Henri Cartan, André Weil, mais également des mécaniciens tels que Henri Villat (1919-1927), qui organisera dès 1920 le 6ème Congrès international des mathématiciens à Strasbourg, et ses élèves, Joseph Pérès qui restera peu de temps à Strasbourg (1920), et René Thiry (1919-1939), ce dernier sera notamment Directeur de l'Institut de Mathématiques et fondateur du Laboratoire de Mécanique des Fluides de Strasbourg. Sachant que Strasbourg sera considérée comme l'antichambre de la Sorbonne. Dans la même période (1919-1939), un certain nombre de physiciens de renom mais souvent plutôt expérimentateurs, seront également acteurs du développement de la mécanique, celle-ci ne pouvant se faire sans des bases solides en mathématiques [3].
*Les dates sont celles de leur séjour strasbourgeois.

## 2     La période germanique (1872-1914)

Le 10 mai 1871, La France et l'Empire allemand, formé en janvier de la même année, signent le traité de Francfort, mettant fin à la guerre. La France doit payer un lourd tribut en cédant l'Alsace et la Moselle, une partie de la Lorraine. Dans ce cadre, l'Université de Strasbourg, université Impériale sous Napoléon, redevient donc en 1872, Impériale mais avec la dénomination germanique « Kaiser Wilhelm Universität-KWU », Strasbourg étant à présent capitale Wilhemienienne et Bismarkienne du Reichsland Alsace –Lorraine, elle le sera jusqu'en 1918. Dès fin 1872, l'Institut de « Philosophie-Sciences Naturelles » est divisée en deux, dont une faculté de « Mathématiques et Sciences naturelles », une nouveauté au sein d'une université prussienne. Kronecker est consulté pour la nomination des mathématiciens, il propose d'abord Friedrich Prym et Elwin Christoffel.
Deux chaires de mathématiques sont alors créées, la première est effectivement confiée à Elwin Bruno Christoffel, venant de Berlin, après un passage à l'École polytechnique fédérale de Zurich, remplaçant Richard Dedekind. La deuxième, en géométrie et mécanique, est finalement attribuée à Theodor Reye, du Polytechnicum d'Aix-la-Chapelle. Theodor Reye est mathématicien mais également féru de météorologie; en 1872 il publie ses études sur les cyclones, et on lui propose de diriger le Centre Océanique de Hambourg, mais il refuse. Tous les deux acceptent facilement les postes proposés, étant de fervents patriotes. Désireux de faire de l'Université strasbourgeoise un établissement d'élite, ils mettent en place un nouvel Institut de mathématiques ainsi que les fameux séminaires de Mathématiques, une institution dans les universités prussiennes [3]. Strasbourg hérite également de l'arrivée de Maximilian Simon, élève de Karl Weierstrass et Ernst Kummer à Berlin (1867), qui se consacrera essentiellement à l'histoire des mathématiques jusqu'en 1912 où il sera professeur honoraire. Christoffel, de santé fragile depuis de longues années, cesse d'enseigner en 1892 et prend sa retraite deux ans plus tard. Son poste laissé vacant est alors confié à Heinrich Weber en 1895.

## 2.1 Heinrich Weber

Heinrich Weber accepte de quitter Göttingen, dont l'université Georg-August est alors considérée comme le centre d'excellence mondial en mathématiques, pour rejoindre Strasbourg essentiellement pour des raisons familiales, sa fille y résidant, mais sans doute aussi pour un meilleur salaire, les taxes étant moins élevées que dans les autres universités prussiennes. Agé de 53 ans, il a déjà un parcours académique exceptionnel. Après des études à Heidelberg, où il suit les enseignements de Gustav Kirchhoff, Otto Hesse, Moritz Cantor et Hermann von Helmholtz, il se rend à Königsberg pour sa thèse d'Etat (1866), dirigée par Franz Ernst Neumann et Friedrich Julius Richelot, ancien élève de Karl Jacobi. Très influencé par les travaux de ce dernier, il enseigne d'abord à Heidelberg (1866-1870), puis remplace Julius Wilhelm Richard Dedekind (1831-1919) à Zurich (1870-1875), puis occupe la chaire de Richelot, décédé, de 1975 à 1883 à Königsberg. Il aura, comme étudiants, David Hilbert ou Hermann Minkowski. Il enseigne ensuite à l'Université technique de Berlin-Charlottenburg, puis à Marburg (1884-1892).

Weber peut être alors déjà être considéré comme un universaliste, comme Dedekind, ses compétences couvrant de nombreux secteurs des mathématiques et de la physique mathématique dont certains aspects de la physique. Dedekind, le dernier doctorant de Carl Gauss à Göttingen (1852), proche de Peter-Gustav Lejeune-Dirichlet nommé pour remplacer Gauss en 1855, est alors en poste à l'Université de Braunschweig (Brunswick) depuis 1862, Weber se lie d'amitié avec Dedekind et ils publient ensemble les œuvres de Bernard Riemann (1976) ainsi qu'un ouvrage sur la théorie des fonctions algébriques d'une variable (1882). Plus précisément, ils développent des méthodes algébriques en géométrie, par le biais des surfaces de Riemann.

Lors de son arrivée à Strasbourg, il rencontre Adolph Krazer (1858-1926), un ancien étudiant de Friedrich Emil Prym (1841-1919) à Würzbourg, nommé professeur extraordinarius (maître de conférences) et Theodor Reye. Ce dernier, depuis son arrivée à Strasbourg, a encadré une douzaine de thèses, ses anciens élèves occupant soit des postes de professeurs de mathématiques, soit d'astronomes, postes souvent confiés à des mathématiciens, Strasbourg possédant un nouvel observatoire depuis 1881.

## 2.2 Contribution de Weber à Strasbourg

En recherches, Weber sera surtout considéré comme algébriste et publiera trois tomes de son « Lehrbuch der Algebra », parus entre 1895 et 1908 pour le troisième volume. A Strasbourg, Weber va également surtout promouvoir et organiser la mise en place des enseignements des futurs professeurs de mathématiques de lycée, c'est-à-dire les nouveaux programmes de la formation des maîtres.  En effet, dès 1898, les nouveaux programmes intègrent une troisième discipline, les mathématiques appliquées (géométrie descriptive, géodésie et mécanique), les deux autres étant les mathématiques pures et la physique.

Weber, avec l'aide de collaborateurs comme Joseph Wellstein (son ancien doctorant en 1894, nommé professeur à Strasbourg en 1904, sur le poste de l'alsacien G. Roth) et son fils Rudolf, d'abord assistant (Privatdozent) à l'Université de Heidelberg, rédige une encyclopédie des mathématiques élémentaires en 3 volumes, destinée aux étudiants et professeurs de lycée.  Le dernier volume auquel son fils est associé, traite des mathématiques appliquées. Deux éditions évolutives et enrichies vont paraître entre 1903 et 1912, et le tome 3 sera consacré à la physique mathématique d'une part, sous la direction de son fils, nommé professeur de physique à l'Université de Rostock, et à des thématiques nouvelles telles que la « graphische Statik », soit les méthodes graphiques d'analyse structurelles utilisées en ingénierie, en calcul des structures et en résistance des matériaux, les probabilités et enfin l'astronomie, cette dernière étant développée par Julius Bauschinger, d'abord directeur de l'Institut de calcul astronomique et professeur d'astronomie théorique à Berlin (1896-1909), puis directeur de l'observatoire de Strasbourg (1909-1919). Concernant la « statique graphique », on pourra citer le livre de Friedrich Schur, mathématicien, spécialiste en géométrie descriptive, également nommé à Strasbourg en 1909 après avoir été recteur de l'Université de Karlsruhe.

Heinrich Weber présidera le troisième Congrès International de Mathématiques à Heidelberg en 1904, à la mémoire de Carl Jacobi, né un siècle avant. Les français Emile Borel, Jules Andrade, Pierre Boutroux, Jacques Hadamard, Paul Painlevé y donneront des conférences, ainsi que d'autres personnalités dont David Hilbert, Hermann Minkowski, Ludwig Prandtl, Arnold Sommerfeld, Strasbourg étant représenté par l'historien Maximilian Simon et Martin Disteli.

La faculté sera ainsi amenée à pourvoir un poste de mathématiques appliquées (sans chaire), ce sera celui d'Adolf Krazer, remplacé d'abord par Disteli (1902), dont les recherches portaient sur les engrenages, celui-ci reviendra ensuite à H. Timerding (1905), élève de Reye et spécialiste de géométrie descriptive, enfin à **Richard Edler von Mises** [4].

## 2.3 Richard Edler von Mises

Sans nul doute, Richard von Mises est méconnu de bien des historiens des sciences qui ne retracent pas l'évolution des mathématiques et de la mécanique des fluides à Strasbourg dans ses débuts. Richard von Mises est né à Lemberg (à présent Lviv en Ukraine) le 19 avril 1883. Ses compétences ont couvert bien des domaines, allant des mathématiques (statistique, théorie des probabilités, géométrie constructive, analyse, équations intégrales et différentielles) aux sciences mécaniques (résistance des matériaux, fluides, aérodynamique, aéronautique).

D'origine austro-hongroise, diplômé ensuite en mathématiques, physique et sciences de l'ingénieur de l'université technique de Vienne, il est nommé assistant du mathématicien Georg Hamel (1877-1954) à Brünn (aujourd'hui Brno) de 1906 à 1909. En 1905, encore étudiant, il publie dans la prestigieuse *Zeitschrift für Mathematik und Physik* (« Revue de Mathématique et de Physique ») un article « Pour une géométrie constructive infinitésimale des courbes planes » (*Zur konstruktiven Infinitesimalgeometrie der ebenen Kurven*).

En 1908, Mises est reçu docteur à l'université de Vienne, avec une thèse sur « La détermination des masses effectives du volant de vilebrequin » (*Die Ermittlung der Schwungmassen im Schubkurbelgetriebe*), puis obtient à Brünn son habilitation à enseigner les sciences de l'ingénieur (dissertation sur la « Théorie des turbines hydrauliques » — *Theorie der Wasserräder*) publié dans le volume 57 de la Zeitschrift für Mathematik und Physik. En 1909, à seulement 26 ans, il est nommé professeur de mathématiques appliquées, en remplacement de Timerding, à l'université Impériale KWU de Strasbourg qui recrute notamment des professeurs parmi l'élite intellectuelle prussienne. On lui offre d'emblée un poste de professeur sans chaire (Extraordinarius), ce qui est rare à cet âge.

Richard von Mises y enseigne alors notamment l'aérodynamique, la construction des aéronefs, dispensant un premier cours universitaire sur le vol à moteur en 1913, ce qui est sans doute une nouveauté mondiale, tout au moins au sein des universités germaniques. Richard von Mises avait l'avantage d'être lui-même pilote et instructeur de vol. C'est également à Strasbourg que von Mises développe ses travaux sur la plasticité et sa théorie de déformation des matériaux, sujet qu'il avait déjà abordé à Brünn et qu'il consolidera occasionnellement plus tard.

Il est l'auteur d'un grand nombre d'ouvrages dont deux furent rédigés en partie pendant son séjour à Strasbourg, un premier sur l'hydromécanique «Elemente der Technische Hydromechanik » paru en 1914, un deuxième, un livre de 320 pages sur l'apprentissage du vol à moteur, « Fluglehre », dont la première édition de 200 pages date de 1918, et mentionne notamment les travaux de Gustave Eiffel, de Ludwig Prandtl et Nicolai Joukowsky.

La Première Guerre mondiale mit fin à ses activités effectives à l'université de Strasbourg. Mises rejoignit l'armée austro-hongroise d'abord comme pilote d'essais, puis comme instructeur de vol pour les officiers et pilotes, et en 1915 dirigea la construction d'un avion de 600 ch (en fait 2 moteurs Daimler de 300 ch chacun entraînant chacun une hélice), le *Mises-Flugzeug*, pour l'armée autrichienne. Terminé en 1916, cet avion de chasse, en fait un bombardier opérationnel avec un profil d'ailes inspiré de ceux développés par Eiffel, ne prit jamais part aux combats, mais von Mises participa comme co-pilote aux premiers vols d'essai en consignant les performances et améliorations à apporter à son premier prototype qui pesait près de 5 tonnes.

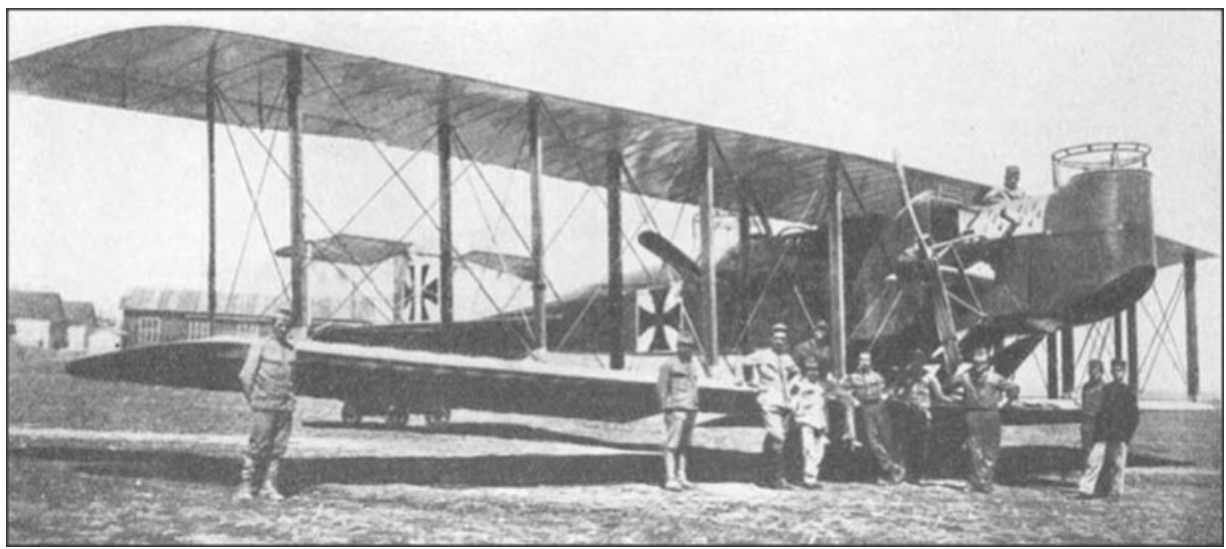

Photo tirée de : von Mises Richard, Ein 600 PS-Großflugzeug vom Jahre 1916, Selected papers, vol. 1, 1963, pp. 530-540, Frank Phillip & et, AMS

**Sa vie après-guerre plus connue.**

En 1920, il fut nommé directeur, avec titre de professeur (Ordinarius), du nouvel Institut des mathématiques appliquées à l'Université de Berlin, après un court passage par les Universités de Francfort, puis Dresde (1919), à la Technische Hochschule, sur un poste de résistance des matériaux, d'hydrodynamique et d'aérodynamique.

Il fonda la célèbre revue *Zeitschrift für Angewandte Mathematik und Mechanik* (« Revue de Mathématique et Mécanique appliquées ») en 1921 et en devint rédacteur en chef, jusqu'à l'arrivée au pouvoir des Nazis en 1933. Il fonda également, en 1922, avec Ludwid Prandtl, la GAMM, "Gesellschaft für Angewandte Mathematik und Mechanik", "Association for Applied Mathematics and Mechanics" qui regroupe actuellement près de 1500 membres. La GAMM est un peu l'équivalent de l'AFM avec une connotation de la Société de mathématiques appliquées et industrielles (SMAI).

Von Mises, d'origine juive, s'exila alors en Turquie et occupa la nouvelle chaire de mathématiques pure et appliquée à l'université d'Istanbul. En 1939, à la mort du président turc Mustafa Kemal Atatürk et face aux troubles politiques menaçants, il émigra aux États-Unis. En 1944, il fut nommé à la chaire Gordon-McKay d'aérodynamique et de mathématique appliquée de l'université Harvard. Richard von

Mises est décédé le 14 juillet 1953 à Boston (MA). IL fonda également, avec Theodore Von Kármán la série d'ouvrage à chapitres « Advances in Applied Mechanics » dont le premier volume date de 1948.

## 3    La période française (1919-1939)

A partir de l'Armistice du 11 novembre 1918, tout va aller très vite même si les autorités allemandes demanderont à leur corps professoral de rester en poste. Le recteur allemand von Thor est révoqué le 30 novembre, et le samedi 7 décembre la KWU(S) – Kaiser-Wilhelms-Universität Strassburg – est dissoute, *deux jours avant l'arrivée triomphale à Strasbourg du Président de la République Poincaré et du président du Conseil Georges Clémenceau* [5]. Les professeurs en place sont congédiés sans pension ni traitement, expulsés avec un minimum de bagages.
Les politiciens et scientifiques parisiens, dont Paul Painlevé et Raymond Poincaré, désiraient d'ailleurs faire de Strasbourg le deuxième pôle universitaire de France après Paris. La faculté des Sciences se met en place ; la mécanique n'existe pas encore, elle reste, comme l'astronomie intégrée aux mathématiques appliquées si elle est théorique, expérimentale, elle sera de plus en plus rattachée à la physique. Le réseau de l'Ecole Normale Supérieure favorise fortement, par le biais de Pierre Weiss, ami d'Aimé Cotton, de Georges Weiss, Paul Appell et Henri Villat, la nomination de jeunes scientifiques issus de l'ENS, en particulier des mathématiciens. Dès janvier 1919, Pierre Weiss (physique), Maurice Fréchet (analyse supérieure), Ernest Esclangon (astronomie), Arnaud Denjoy (mathématiques), Edmond Rothé (physique du globe), Gustave Ribaud (physique) sont recrutés ; peu après c'est **Henri Villat** en mécanique rationnelle qui rejoint Strasbourg, de même que **René Thiry** ou André Danjon (astronome) quelque temps plus tard fin 1919. Arnaud Denjoy, bien que nommé professeur de mathématiques générales à la faculté des sciences de l'université de Strasbourg, est alors surtout maintenu en mission à Utrecht. Edmond Bauer, après son séjour à Zurich et son retour auprès de Jean Perrin à la Faculté des Sciences de Paris, est nommé sur un poste de maître de conférences en physique mathématique (le seul poste alors de ce type en province). Enfin **Joseph Pérès** obtient après Toulouse (1916) un poste de maître de conférences de mathématiques générales à la faculté des Sciences de Strasbourg (1919) avant de rejoindre Marseille en 1921. Georges Valiron, ancien doctorant d'Emile Borel (1914), occupera la chaire de professeur de calcul différentiel et intégral à partir de 1919. Dans le cadre de ces nominations rapides, il s'agissait en particulier de mettre en place, dès la rentrée universitaire de 1919, les enseignements scientifiques analogues à ceux dispensés par ailleurs en France, notamment à Paris [6].

### 3.1 Mise en place de l'Institut de Mathématiques – Enseignement

Maurice Fréchet prend la direction de l'Institut de Mathématiques et le programme d'enseignement est rapidement constitué [7]. En 1921, les acteurs principaux, en « Mécanique » sont Pérès, Villat et Thiry, l'astronome Alexandre Véronnet [8] et le physicien Edmond Bauer. Dans un premier temps, pour résumer le séjour et la carrière ultérieure des trois premiers, on peut se contenter des données et dates suivantes, extraites des Archives Départementales du Bas-Rhin (ADBR):

- *Joseph Pérès sera maître de conférences de mathématiques générales de 1919 à 1921, avant son départ à Marseille sur un poste de professeur de mécanique rationnelle en 1921, puis sa nomination en 1932 à la faculté des Sciences de Paris sur la chaire de mécanique rationnelle (1942-1950), puis sur la chaire de mécanique des fluides et applications (1950-1962) et doyen de la faculté des Sciences de Paris (1954-1961).*

- *Henri Villat 1919-1927 occupe la chaire de mécanique rationnelle de 1919 à 1927, avant son départ pour la Sorbonne comme professeur de mécanique des fluides en avril 1927.*
- *René Thiry, élève de Henri Villat est chargé de cours en mathématiques générales (1921-1925), puis maître de conférences en mécanique des fluides (1925-1939), chaire de mécanique des fluides créée en 1925, et prendra la direction de l'Institut de Mécanique des Fluides de Marseille en octobre 1939.*

L'astronome Alexandre Véronnet [8], spécialiste de cosmologie et de géodésie, arrivé au printemps 1919 pour travailler à l'Observatoire de Strasbourg, rejoint la nouvelle équipe des astronomes strasbourgeois, Ernest Esclangon, nommé professeur d'astronomie, et André Danjon, nommé astronome-adjoint ; André Danjon, plus jeune que Véronnet, ayant été recommandé par Emile Borel à Esclangon. Il va d'abord collaborer avec Henri Villat, en qui il trouve un soutien sincère, qui lui confie des enseignements en mécanique rationnelle. Villat assure alors 2h de cours hebdomadaires et Véronnet est chargé de 2h de conférences, l'enseignement de l'astronomie étant confiée à Ernest Esclangon et André Danjon, les mathématiques sont confiées à Fréchet, Valiron, Pérès, Arnaud Denjoy (cependant en mission à Utrecht), René Thiry sollicité par Villat et Louis Antoine, rendu aveugle pendant la guerre. La physique mathématique est confiée à Edmond Bauer, ancien assistant de Jean Perrin et collaborateur de Paul Langevin et Marie Curie, avant la guerre, nommé maître de conférences et rattaché au grand laboratoire de magnétisme de Pierre Weiss de la Faculté des Sciences.

## 3.2 Henri Villat

En 1920, Henri Villat organise lui-même **le 6ème Congrès international des mathématiciens, le 5ème s'était tenu à Cambridge en août 1912, du 22 au 30 septembre** à Strasbourg, avec comme secrétaire René Thiry, alors que d'autres mathématiciens strasbourgeois dont Maurice Fréchet s'y sentent moins impliqués ; les mathématiciens des nations vaincues dont les brillants ressortissants allemands y étant exclus, dont Richard von Mises ou Theodore von Kármán. Cela sera également le cas en 1924 à Toronto. Les conférences générales seront notamment données par Sir Joseph Larmor (Cambridge), Léonard .E. Dikson (Chicago), Charles-Jean de la Vallée Poussin (Louvain), Vito Volterra (Rome), Niels Erik Nörlund (Lund). Présidé par Emile Picard, Camille Jordan étant président d'Honneur, le congrès réunira 200 scientifiques de 27 nations, mathématiciens, astronomes dont certains sont également connus pour leurs travaux en physique ou en mécanique tels que Alfred Greenhill, Stanisław Zaremba, Marcel Brillouin, Albert Châtelet, Arnaud Denjoy, Georges Valiron, Gabriel Koenigs ou encore Dimitri Riabouchinsky qui aura le soutien de Paul Appell, d'Emile Picard, Gabriel Koenigs, Paul Painlevé, Henri Villat, Henri Bénard entre autres.

En parallèle, suite au Congrès de Strasbourg, Henri Villat devient rédacteur puis directeur du Journal de mathématiques pures et appliquées (1921-1970) fondé par Joseph Liouville en 1836, que lui confient Emile Picard et Camille Jordan, reprend les Nouvelles annales de mathématiques (octobre 1922-1928), en collaboration avec Joseph Pérès et Raoul Bricard, et crée et dirige deux collections, le Mémorial des sciences mathématiques (1925-1972) d'une part, et le Mémorial des sciences physiques (1928-1963) avec comme co-rédacteurs Jean Villey, puis Charles Fabry, enfin Gustave Ribaud.

L'Institut de Mathématiques de Strasbourg organise en 1922 une série de cours spéciaux destinés à encourager la recherche scientifique. A côté de Fréchet, Valiron et Thiry, Villat propose sur un semestre un cours « de développements sur la mécanique analytique et quelques récentes applications ».

Les années suivantes, Henri Villat donnera des cours sur les théories des tourbillons (1923), recherches par certaines généralisations de l'équation différentielle de Lamé et sur la théorie des surfaces minima (1924), mouvements des fluides peu visqueux (1925), théorie des ondes liquides (1926). Ces leçons étant un prélude à ses futurs ouvrages.

Dès 1925, Henri Villat séjourne souvent à Paris, pour seconder ou remplacer dans ses enseignements, Paul Painlevé, titulaire de la chaire de « Mécanique des fluides et applications » créée à la faculté des sciences de Paris (Sorbonne) en 1923 mais très occupé par ses tâches ministérielles – ministre de la Guerre -, dont Villat est le « protégé ». Il sera alors amené à donner deux séries de cinq conférences (avril et mai 1925) et à rédiger des leçons sur « la théorie des sillages en hydrodynamique », ainsi que des leçons **sur la Mécanique des Fluides, en collaboration avec Paul Painlevé et Charles Camichel, 1925**), puis de quatre séries de conférences sur la mécanique des fluides (avril à juin 1926), toujours à la Faculté des sciences de Paris.

Henri Villat dirigera, à Strasbourg, trois thèses, celle de René Thiry, Maurice Roy et Jacques Bergé et participera également à deux autres thèses :

- celle de René Thiry (6 juillet 1921) Doctorat ès Sciences « *Sur les solutions multiples des problèmes d'hydrodynamique relatifs aux mouvements glissants* » (Jury : Brillouin, Villat, Esclangon)
- celle de Maurice Roy (7 avril 1923), Doctorat ès Sciences « *Recherches sur les surfaces portantes en aérodynamique. La théorie de Prandtl* », Jury Esclangon, Villat, Thiry)
- *celle de Jacques Berge (initialement prévue en juin 1925, soutenue le 10 mai 1926)* Doctorat ès Sciences Mathématiques « *Quelques applications de l'hydrodynamique »* Jury Esclangon, Villat, Thiry
- **On peut ajouter une première thèse d'un confrère mathématicien que Henri Villat a bien connu : Louis Antoine, grand blessé de guerre,** *Sur l'homéomorphie de deux figures et de leurs voisinages (9 juillet 1921,Strasbourg, jury : Président Henri Lebesgue, Villat, Fréchet, Valiron)*  Louis Antoine, maitre de conférences de Mathématiques à Strasbourg, va, à sa demande, intégrer l'Université de Rennes dès 1922, pour convenances personnelles
- De même, Henri Villat va co-encadrer, avec Georgios Remoundos, professeur à l'Université d'Athènes, et ancien doctorant d'Emile Picard, la thèse d'analyse et d'algèbre supérieure de Spyridion Sarantpoulos (mars 1923).

**Remarque : Un mot sur Paul Flamant**

A l'Institut de Mathématiques, auquel Henri Villat est rattaché, une autre thèse ès sciences mathématiques sera soutenue en 1924, celle de Paul Flamant, ancien élève de l'ENS, alors chargé de cours à la Faculté des Sciences de Strasbourg, devant la commission d'examen composée d'Emile Borel (Président), M. Fréchet, E. Esclangon et G. Valiron. En fait, Paul Flamant (1892–1940), entré premier à l'ENS en 1913, sera sous-lieutenant de l'armée française, blessé et emprisonné quatre ans au camp de Charleroi. Reçu premier à l'agrégation en 1919, il est préparateur à l'ENS jusqu'en 1921, année où il revient Strasbourg comme chargé de cours de mathématiques à la faculté des Sciences, ce dans la chaire d'Arnaud Denjoy, en mission à Utrecht. En 1924, soutenant sa thèse, il rédige également les notes du cours Peccot de Gaston Julia au Collège de France de 1920 dans la monographie *Leçons sur les Fonctions Uniformes à Point Singulier Essentiel Isolé* (1924). En 1926, il est nommé maître de conférences pour remplacer René Thiry et enseigne la mécanique rationnelle en 1928 à Clermont-Ferrand, et est nommé professeur de mécanique rationnelle la même année dans cette ville. Il retourne à Strasbourg comme professeur de mathématiques générales à la faculté des Sciences de Strasbourg en 1929, puis dans la chaire de calcul différentiel en 1935.

## 3.3 René Thiry

René Thiry est professeur au Centre d'études militaires de Strasbourg pendant quelques mois (avril-septembre 1919), où il côtoie Henri Villat; d'octobre à novembre 1919, il est professeur de mathématiques spéciales au Lycée Kléber de Strasbourg. Dès novembre 1919, il est chargé de cours de mathématiques à la faculté des Sciences de Strasbourg, en suppléant Arnaud Denjoy, nommé professeur de mathématiques générales à la faculté des Sciences de Strasbourg, mais toujours maintenu en mission à l'université d'Utrecht sur une chaire de théorie des fonctions. René Thiry prend en 1921 le poste de maître de conférences devenu vacant suite au départ de Joseph Pérès pour Marseille, Paul Flamant étant désigné pour le remplacer sur le poste de Denjoy.

Au premier semestre 1922, René Thiry est chargé au collège de France, au titre de la fondation Peccot, de 30 leçons portant sur les recherches modernes sur la résistance des fluides.

En 1925, Thiry passe du jury d'admission de l'Ecole normale supérieure à celui de l'Ecole navale. Il donne en 1929, des conférences bénévoles destinées à l'agrégation féminine, Edmond Rothé, directeur de l'Institut de Physique du Globe devenant doyen de la faculté des Sciences. Charles Sadron fait alors une communication à la SFP sur l'étude des tourbillons annulaires, les travaux étant issus de son séjour comme étudiant de Georges Bouligand à Poitiers.

Esclangon est appelé en 1930 à la direction des observatoires de Paris et Meudon, et est élu membre de l'Académie des Sciences. Grâce à des appuis divers obtenus ou escomptés, dont celui la chambre du Commerce, Thiry demande la création d'un enseignement de mécanique des fluides et l'installation d'un laboratoire technique d'application, la faculté des Sciences espérant être pourvue de la création d'une chaire théorique de mécanique des Fluides. Dans cette même période, Henri Cartan est nommé chargé de cours à la faculté des sciences puis maître de conférences (1931).

En 1932, Georges Valiron, directeur de l'Institut de Mathématiques, est chargé d'une suppléance à la Sorbonne. Il est alors remplacé par René Thiry à la direction de cet Institut, et Henri Cartan assure son enseignement. Le ministère de l'Air ayant continué de verser à la faculté ses subventions, l'enseignement de la mécanique des fluides a pu être complété cette année encore par les conférences de Charles Sadron qui a pris un congé au Lycée Kléber et s'est consacré à l'aérodynamique (Rapport de l'Université, 1932). Cette même année, au mois de mai, Georges Bouligand, professeur à l'Université de Poitiers dans la chaire de calcul intégral et différentiel après celle de mécanique rationnelle, en visite à Strasbourg, donne trois conférences « *très appréciées sur ses belles recherches de géométrie infinitésimale* ». N'oublions pas que le DES de Charles Sadron était supervisé par Georges Bouligand, et que les travaux permettront à Sadron de publier son premier article en 1926 [9], Georges Bouligand l'ayant explicitement cité dans un Compte-rendu de 1925.

**En 1932, la Faculté des Sciences vient enfin de créer son Laboratoire de mécanique des Fluides**, René Thiry en prend la direction en plus de celle de l'Institut de Mathématiques, ce qui représente une lourde charge. C'est également l'année où et Charles Sadron, collaborateur du Ministère de l'Air, et Louis Néel, assistant depuis 1928, soutiennent leur doctorat ès sciences physiques au sein du Laboratoire de Pierre Weiss, Sadron sur les moments ferromagnétiques des éléments et le système périodique le 7 janvier, Néel sur les propriétés magnétiques des corps le 11 décembre. Le jury de Sadron sera composé de Pierre Weiss, René Thiry (Mécanique rationnelle) et Gustave Ribaud (Physique expérimentale). Celui de Néel de Pierre Weiss, Héloïs Ollivier (Physique générale), Lucien Hackspill (Chimie minérale) et Paul Soleillet (Physique mathématique).

Avec ses collègues mathématiciens Georges Cerf, Henri Milloux et Paul Flamant, René Thiry prend part, pendant les « vacances » de septembre 1932, aux travaux du Congrès international des mathématiciens à Zurich, Elie Cartan y représentant officiellement la France. André Weil est nommé maître de conférences de mathématiques en 1933 en remplacement de Georges Valiron, parti pour la Sorbonne et l'ENS.

André Danjon est nommé doyen de la Faculté des Sciences en date du 18 août 1935, et René Thiry son assesseur par arrêté du 9 décembre 1935. Henri Cartan est nommé professeur titulaire de la chaire de Mathématiques générales le 1er janvier 1936 et c'est André Weil qui le remplace sur le poste de maître de conférence au 1er avril. Le groupe Bourbaki se met en place en juillet 1935 avec Henri Cartan, son ami André Weil, ainsi que Jean Dieudonné, Szolem Mandelbrojt, Claude Chevalley, René de Possel, Charles Ehresmann et Jean Delsarte.

Charles Sadron revient alors du laboratoire d'aérodynamique de chez von Karman au CalTech en juillet 1934, bien que von Karman lui ait demandé de rester. Néanmoins une amitié profonde les liera par la suite. Sadron publie ses travaux sur les lois de frottement en conduites lisses et rugueuses et travaille à nouveau dans le laboratoire de René Thiry en temps que collaborateur scientifique du Ministère de l'Air.

René Thiry va superviser le doctorat ès Sciences physiques de Louis-André Sackmann (6 juillet 1936), l'aide matérielle venant du Service de Recherches de l'Aéronautique du Ministère de l'Air. Sa thèse plutôt expérimentale, la première thèse de mécanique des fluides générale, porte sur « L'écoulement des fluides au voisinage des points singuliers des obstacles », la proposition de la 2ème thèse par la Faculté des Sciences concernera « Les lois générales de la viscosité des liquides ». Le jury est présidé par René Thiry, chaire de Mécanique rationnelle, les examinateurs sont Gabriel Foëx, chaire de Physique expérimentale et Henri Weiss, chaire de Physico-Chimie du Pétrole.

1937 à 1939 seront des années charnières. Le 1er novembre 1937, Louis Néel est nommé professeur de physique générale, en remplacement de Pierre Weiss admis à valoir ses droits à la retraite, Charles Sadron prend le poste de Louis Néel, la maîtrise de conférences de physique mathématique, qu'occupait Paul Soleillet, nommé professeur titulaire à la Faculté des Sciences de Poitiers, est confiée à Jacques Yvon. Charles Sadron s'investit dans la biréfringence d'écoulement dès 1936, quittant progressivement le laboratoire de mécanique des fluides de René Thiry.

**Par décret du 3 mars 1939, René Thiry est transféré dans la chaire de mécanique des fluides de la Faculté des Sciences de Marseille en octobre 1939**. André Weil est nommé dans la chaire laissée vacante de mécanique rationnelle, mais revenu de Princeton, il part en Finlande, fuyant la proche guerre. André Weil ne reviendra pas sur Strasbourg au moment de la déclaration de la guerre, ce qui lui vaudra de multiples ennuis ; et Charles Ehresmann, chargé de recherches au CNRS, obtient la maîtrise de mathématiques.

## 3.4 Autres contributions venant des mathématiciens et physiciens

En dehors de celle Charles Sadron, plusieurs contributions au développement de la mécanique viendront des physiciens de Strasbourg, celle ponctuelle de Serge Nikitine, des chercheurs de l'Ecole Nationale Supérieure du pétrole et des Combustibles Liquides en rhéologie, de l'Institut de Physique du Globe en météorologie notamment (Edmond Rothé), et des astronomes de l'Observatoire des Strasbourg. Plusieurs scientifiques, avant d'être nommés à Strasbourg, seront mobilisés à la commission d'artillerie navale de Gâvres où se trouve un champ d'essai balistique (arsenal de Lorient), dit polygone, unique centre d'essai d'artillerie de la Marine. Ainsi le physicien Gabriel Foëx poursuivra également des recherches sur les tirs d'artillerie et les phénomènes de détonation en collaboration avec **Joseph Kampé de Fériet (1925-1927)**, La rencontre entre Kampé de Fériet, d'abord mathématicien, et Foëx, spécialiste de la physique expérimentale, sera très bénéfique pour les futurs axes de recherches et compétences de Kampé de Fériet tout au long de sa brillante carrière au sein de l'Institut de Mécanique des Fluides de Lille. Cette commission aura également intégré plusieurs mathématiciens (Jules Haag, Arnaud Denjoy, Georges Valiron, Ernest Esclangon) mobilisés dès 1915, ceux-ci mettront également en valeur leurs recherches en balistique (Valiron, Esclangon), entre les deux guerres. En 1925, Esclangon, promoteur du repérage des canons par le son, publiera un ouvrage de 388 pages sur l'acoustique des canons et des projectiles.

## Conclusion

On l'aura compris. Après une domination germanique, où les mathématiciens allemands vont marquer de leur emprunte le paysage universitaire de Strasbourg, le gouvernement français donnera tous les moyens à l'Université de Strasbourg pour se redresser et rayonner. Lors de la déclaration de la guerre début septembre 1939, l'Université de Strasbourg aura deux jours pour se replier sur Clermont-Ferrand, de même que la population strasbourgeoise qui va s'exiler rapidement dans « la France de l'intérieur », laissant une ville fantôme. Mais ceci est un autre pan de l'histoire qui marquera la période 1939-1945, l'après-guerre, la reconstruction de l'Université de Strasbourg Française. Les jalons seront posés pour confirmer la notoriété strasbourgeoise en Mathématiques après la deuxième guerre mondiale avec le groupe Bourbaki, René Thom, André Lichnérowicz (1941-1949), l'Institut de Recherche Mathématique Avancée (IRMA), Institut devenu, le 1er janvier 1966, le premier laboratoire associé au CNRS (L.A. n° 1, puis URA 001), sous la houlette de Jean Frenkel et Georges Reeb, et en Mécanique avec la naissance d'un Institut de Mécanique des Fluides en 1970 grâce aux efforts constants de Louis Sackmann pour obtenir de grands locaux (Institut devenant Institut de Mécanique des Fluides **et des Solides** ensuite en 2000, issu du laboratoire de René Thiry) ou encore en physico-chimie des polymères avec ce qui deviendra l'Institut Charles Sadron (d'abord Centre d'Études de la Physique des Macromolécules – CEPM, premier laboratoire propre du CNRS en province en 1947, puis Centre de Recherche sur les Macromolécules – CRM, unité associée à l'Université de Strasbourg, 1954).

## Références